\documentclass{article}

\PassOptionsToPackage{numbers, compress}{natbib}

\usepackage[final]{neurips_2023_ml4ps}

\usepackage[utf8]{inputenc} 
\usepackage[T1]{fontenc}    
\usepackage{ae,aecompl}

\usepackage{hyperref}       
\usepackage{url}            
\usepackage{booktabs}       
\usepackage{amsfonts}       
\usepackage{nicefrac}       
\usepackage{microtype}      
\usepackage{hyperref}

\usepackage{graphicx}	
\usepackage{amsmath}	
\usepackage{amssymb}	
\usepackage{algorithm}
\usepackage{algorithmic}
\usepackage{caption}
\usepackage{subcaption} 

\title{deep-REMAP: Parameterization of Stellar Spectra Using Regularized Multi-Task Learning}

\author{%
  Sankalp Gilda\thanks{\url{www.linkedin.com/in/sankalp-gilda}} \\
  Machine Learning Collective \\
   \texttt{sankalp.gilda@gmail.com} \\
}

\begin{document}

\maketitle

\begin{abstract}
Traditional spectral analysis methods are increasingly challenged by the exploding volumes of data produced by contemporary astronomical surveys. In response, we develop deep-\underline{R}egularized \underline{E}nsemble-based \underline{M}ulti-task Learning with \underline{A}symmetric Loss for \underline{P}robabilistic Inference ($\rm{deep-REMAP}$), a novel framework that utilizes the rich synthetic spectra from the PHOENIX library and observational data from the MARVELS survey to accurately predict stellar atmospheric parameters. By harnessing advanced machine learning techniques, including multi-task learning and an innovative asymmetric loss function, $\rm{deep-REMAP}$ demonstrates superior predictive capabilities in determining effective temperature, surface gravity, and metallicity from observed spectra. Our results reveal the framework's effectiveness in extending to other stellar libraries and properties, paving the way for more sophisticated and automated techniques in stellar characterization.

\end{abstract}

\section{Introduction}\label{sec:introduction}
\label{sec:introduction}
Advancements in computers, telescope designs, and funding have revolutionized astronomy, leading to expanded survey volumes and resolution \citep{york2000sloan, eisenstein2011sdss}. These developments have enabled surveys like SDSS, SEGUE \citep{segue}, RAVE \citep{rave}, BOSS \citep{boss}, and LAMOST \citep{lamost} to collect spectra for millions of stars, creating a data-rich environment for cutting-edge research. Complementing this, the PHOENIX library has provided a comprehensive set of synthetic spectra, essential for developing and testing new spectral analysis techniques. The latest data releases from Gaia-ESO Survey \citep{gaiaeso}, DESI \citep{desi}, and LSST \citep{lsst} are set to further expand this corpus, highlighting the necessity for advanced computational methods capable of handling such vast and complex datasets. The increasingly prominent role of Machine Learning (ML), particularly Deep Learning (DL), in processing and analyzing these data reflects this need, with techniques such as convolutional neural networks (CNNs) and recurrent neural networks (RNNs) proving effective for tasks like modeling SEDs of galaxies \citep{mirkwood}, improving image quality at telescopes \citep{cfht, neurips_cfht}, stellar spectrum analysis \citep{deepremap_abstract_aas} and inferring redshifts from gamma ray bursts \citep{gamma_ray}.

Traditional spectral analysis methods, though successful in the past, now face significant challenges due to the surge in data volume and the diversity of spectral resolutions provided by these surveys. High-resolution techniques like the Equivalent Widths (EW) Method \citep{valenti2005spectroscopic, equivalent_width} are less suited for the moderate-resolution spectra that are now increasingly common.  Spectral synthesis, an alternative, has shortcomings including dependency on atomic line databases \citep{ghezzi2014accurate}.

\begin{figure*}
\centering
\includegraphics[width=1\linewidth]{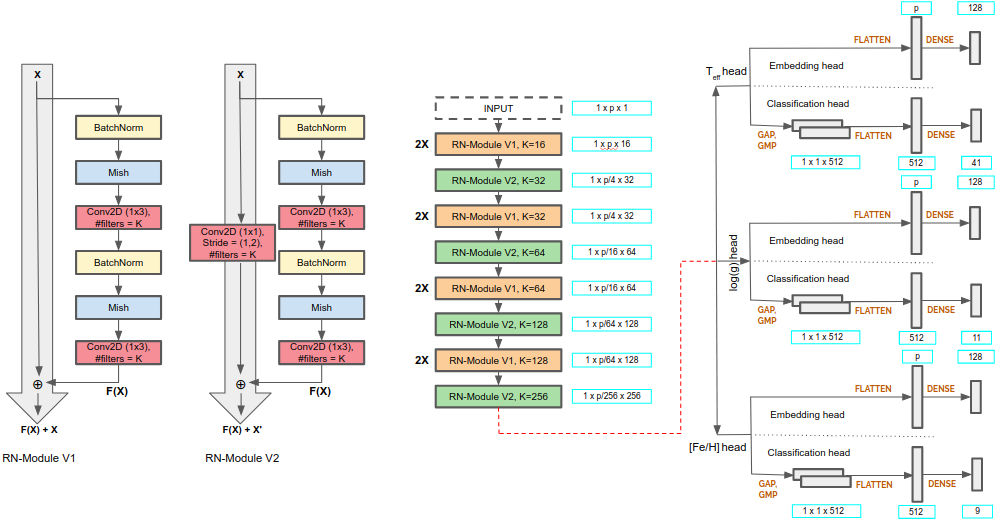}
\caption{Schematic representation of the $\rm{deep-REMAP}$ neural network architecture highlighting the input layer, convolutional layers, pooling layers, dense layers, and the two heads for of the three stellar parameters.}
\label{fig:deep_remap}
\end{figure*}

Recent years have seen the development of ML algorithms for stellar parameterization, from multi-layer perceptrons \citep{mlp_1,mlp_2,mlp_3} to convolutional neural networks (CNNs) like ``The Cannon'' \citep{cannon}, ``The Payne'' \citep{payne}, and ``StarNet'' \citep{starnet}. However, these works either use simple networks, require significant labelled data, or lack generalization to out-of-sample distributions. Our proposed framework, $\rm{deep-REMAP}$ (see Figure \ref{fig:deep_remap}), is designed to leverage the synergy between the synthetic spectra from the PHOENIX library and the observational data from the MARVELS survey. It addresses these contemporary challenges by employing a multi-task learning approach, enabling the simultaneous prediction of multiple atmospheric parameters. We apply it to MARVELS survey spectra \citep{marvels_ge_intro}, validate it on stars with known parameters, and extract stellar atmospheric parameters for 732 FGK giant star candidates.

$\rm{deep-REMAP}$'s architecture, which integrates recent advancements in regularization and loss function design, is particularly well-suited to the complex task of extracting stellar characteristics from the vast and varied data produced by today's spectroscopic surveys.

\section{Data and Pre-Processing}\label{sec:data}
\subsection{MARVELS Spectra}\label{sec:marvels spectra}
We leverage the rich dataset from SDSS-III's Multi-object APO Radial Velocity Exoplanet Large-area Survey (MARVELS), featuring spectra collected from 5,500 stars using a medium-resolution spectrograph at Apache Point Observatory \citep{marvels_ge_intro, marvels_telescope, marvels_ge_2}. Prior analyses have scrutinized the UF2D and UF1D data processing pipelines \citep{grieves2017exploring}, along with the meticulous target selection process \citep{Paegert_2015}. 

Expanding upon the spectral indices method \citep{ghezzi2014accurate}, \citep{grieves2018chemo} refined the derivation of stellar parameters for all observed stars. Following quality control and uniqueness checks, as well as parameter filtering, we narrowed the dataset down to 2,343 dwarfs, whose stellar parameters have subsequently been used for the fine-tuning of our network. The validation process involved 30 calibration stars, and the network was also tasked with predicting parameters for the remaining 732 giant and sub-giant stars (Nolan Grieves, private communication).

The pre-processing of MARVELS spectra entailed meticulous normalization, continuum removal, and a novel approach to cosmic ray mitigation. These steps were critical in ensuring the quality of the dataset, which is paramount when employing deep learning models for spectral analysis. The curated dataset enables the $\rm{deep-REMAP}$ framework to effectively learn the intricate patterns and nuances within the spectra, facilitating a more accurate prediction of stellar atmospheric parameters.



\subsection{PHOENIX Spectra}\label{sec:phoenix_spectra}
The PHOENIX library offers an extensive collection of high-resolution synthetic spectra, pivotal for the calibration and benchmarking of spectral analysis algorithms. Utilizing the state-of-the-art synthetic modeling techniques, the library provides coverage over a wide range of stellar parameters, making it an indispensable resource for this study. For the purposes of $\rm{deep-REMAP}$, we selected a subset of the PHOENIX spectra that closely matches the parameter space of the MARVELS survey to ensure a consistent and comprehensive training regimen for our model. Since MARVELS stars are expected to be F,G and K type stars, we limit the PHOENIX stellar parameter to: 4,850 $\leq\lambda\leq$ 5,750 \AA, 3,000 $\leq$ T$_{\rm{eff}}$ $\leq$ 7,000 K, 1.0 $\leq \log\;\rm{g} \leq$ 5.5, -2 $\leq$ [Fe/H] $\leq$ 1. Below, we describe how the synthetic PHOENIX spectra are brought in line with MARVELS spectra to reduce the discrepancies between them, thus allowing for successful transfer learning.

Incorporating these synthetic spectra allows for the fine-tuning of $\rm{deep-REMAP}$'s predictive capabilities, providing the network with robust training data that encapsulates the theoretical variance across different stellar atmospheres. This integration is particularly beneficial for the recognition of subtle spectral features that may not be as prominent in observational data, thereby enhancing the model's generalizability and accuracy in parameter estimation.


\subsection{Pre-processing and Data Augmentation}\label{sec:preprocessing}
Synthetic spectra often require calibration to align with observed spectra, a process termed traversing the \emph{synthetic gap}. To utilize the PHOENIX synthetic grid for the MARVELS spectra, we perform several pre-processing and data augmentation steps:

\begin{enumerate}
    \item \textbf{Down-convolution}: PHOENIX spectra are convolved with Gaussian kernels to lower resolution, accommodating resolution variation in the MARVELS spectra.
    \item \textbf{Resampling}: PHOENIX spectra are resampled onto the MARVELS wavelength grid via cubic spline interpolation.
    \item \textbf{Noise Addition}: Gaussian noise is added to PHOENIX spectra, reflecting the Signal-to-Noise (SNR) distribution in MARVELS spectra.
    \item \textbf{Continuum Normalization}: Continuum is removed from both datasets using a uniform method, ensuring consistency.
\end{enumerate}

Steps 1 and 3 also serve as data augmentation, enhancing the training set size and model robustness. For step 4, we develop a novel continuum normalization routine (see Algorithm \ref{algo:cont_norm}), superior to existing methods in tests on synthetic spectra. In future we plan to compare this with advanced algorithms in literature \citep{cont_norm_wavelet, cont_norm_framelet, cont_norm_wavelet_kalman}. 

MARVELS spectra required careful pre-processing to accurately estimate their continuum levels, which included removal of false features \citep{grieves2017exploring, grieves2018chemo}, and application of the continuum finding routine. Despite varying SNRs in the MARVELS spectra, our method effectively managed potential misestimations. Concluding these steps, we have 100,000 processed PHOENIX spectra, and 3,075 processed MARVELS spectra.

\section{Implementation}\label{sec:implementation}
We design the $\rm{deep-REMAP}$ architecture to accurately predict stellar parameters by discretizing the continuous spectral labels into discrete classes. We demonstrate this using the effective temperature (T$_{\rm{eff}}$) -- we divide the range into bins and model these bins using Gaussian distributions, which reflects the measurement uncertainties inherent in T$_{\rm{eff}}$ estimates and is expressed by the equation:

\begin{equation}
p(T_{\rm{eff}}|x) = \frac{1}{\sqrt{2\pi\sigma^2}}\exp\left(-\frac{(T_{\rm{eff}}-\mu)^2}{2\sigma^2}\right),
\end{equation}

where \( x \) is the input spectrum, \( T_{\rm{eff}} \) is the effective temperature, \( \mu \) is the mean value for the bin, and \( \sigma \) is the standard deviation.

We apply asymmetric label smoothing, a recent advancement in training classification models, to mitigate the impact of these uncertainties, thereby forming a cost vector that guides the neural network towards more reliable predictions. We introduce triplet loss to our training process, a significant improvement over traditional loss functions. This loss function helps in embedding the parameters in a space that enhances the model's predictive accuracy by:

\begin{equation}
\mathcal{L}_{\rm{triplet}} = \max \left( d(a, p) - d(a, n) + \text{margin}, 0 \right),
\end{equation}

where \( d(\cdot) \) denotes a distance metric, \( a \) is an anchor input, \( p \) is a positive input of the same class as the anchor, and \( n \) is a negative input of a different class.

Our final loss function, \( L_{\rm{final}} \), combines cross-entropy loss and triplet loss to balance classification accuracy with embedding effectiveness:

\begin{align}\label{eqn:loss_final}
    L_{final} &= L_{\rm{T}_{\rm{eff}},\rm{focal}} + \lambda_{\rm{T}_{\rm{eff}}} L_{\rm{T}_{\rm{eff}},\rm{triplet}}\\ \nonumber
    &+ L_{\log\;\rm{g},\rm{focal}} + \lambda_{\log\;\rm{g}} L_{\log\;\rm{g},\rm{triplet}}\\ \nonumber
    &+ L_{\rm{[Fe/H]},\rm{focal}} + \lambda_{\rm{[Fe/H]}} L_{\rm{[Fe/H]},\rm{triplet}} 
\end{align}

with \( \lambda \)s being weights that balance the two loss components. See Fig. \ref{fig:triplet_val}) for the results on the validation set. The three $\lambda$s are empirically chosen to be 0.001.

The learning rates for the fine-tuning phase undergo a gradual reduction, thus preserving the knowledge in the shallower layers while facilitating faster modifications in the deeper layers. The number of epochs per cycle and the number of cycles are determined empirically. The weights of the model are set to the running average of the weights recorded at the end of each cycle, a technique known as Stochastic Weight Averaging (SWA). Lastly, the model's performance is evaluated on the stellar parameters of 30 calibration stars from MARVELS.

We determine the optimal model complexity through an exhaustive 10-fold cross-validation, selecting the appropriate number of residual blocks and filters to minimize \( L_{\rm{final}} \). This minimization ensures that our model generalizes well and remains robust against overfitting. Through this strategic implementation, we position $\rm{deep-REMAP}$ at the forefront of modern machine learning applications in spectral analysis, showcasing its prowess in automated stellar parameter inference.

\section{Results}\label{sec:results}
We applied the $\rm{deep-REMAP}$ framework to analyze a cohort of 732 FGK giant star candidates, extracting three key parameters: \( T_{\text{eff}} \), [Fe/H], and \( \log(g) \). Our results confirm that approximately 80\% of these candidates are FGK giants, as their \( 3\sigma \) parameter values align with the expected ranges for giants ( \( 4000 \, \text{K} \leq T_{\text{eff}} \leq 6000 \, \text{K} \), \( -0.5 \, \leq \text{[Fe/H]} \leq 0.3 \), and \( 0.0 \leq \log(g) \leq 3.0 \)). The remaining 20\% were re-classified as FGK dwarfs, with their parameters falling within the dwarf ranges ( \( 5000 \, \text{K} \leq T_{\text{eff}} \leq 7000 \, \text{K} \), \( -0.3 \, \leq \text{[Fe/H]} \leq 0.5 \), and \( 3.5 \leq \log(g) \leq 5.0 \)).

\begin{figure*}
    \centering
    \includegraphics[width=1\textwidth]{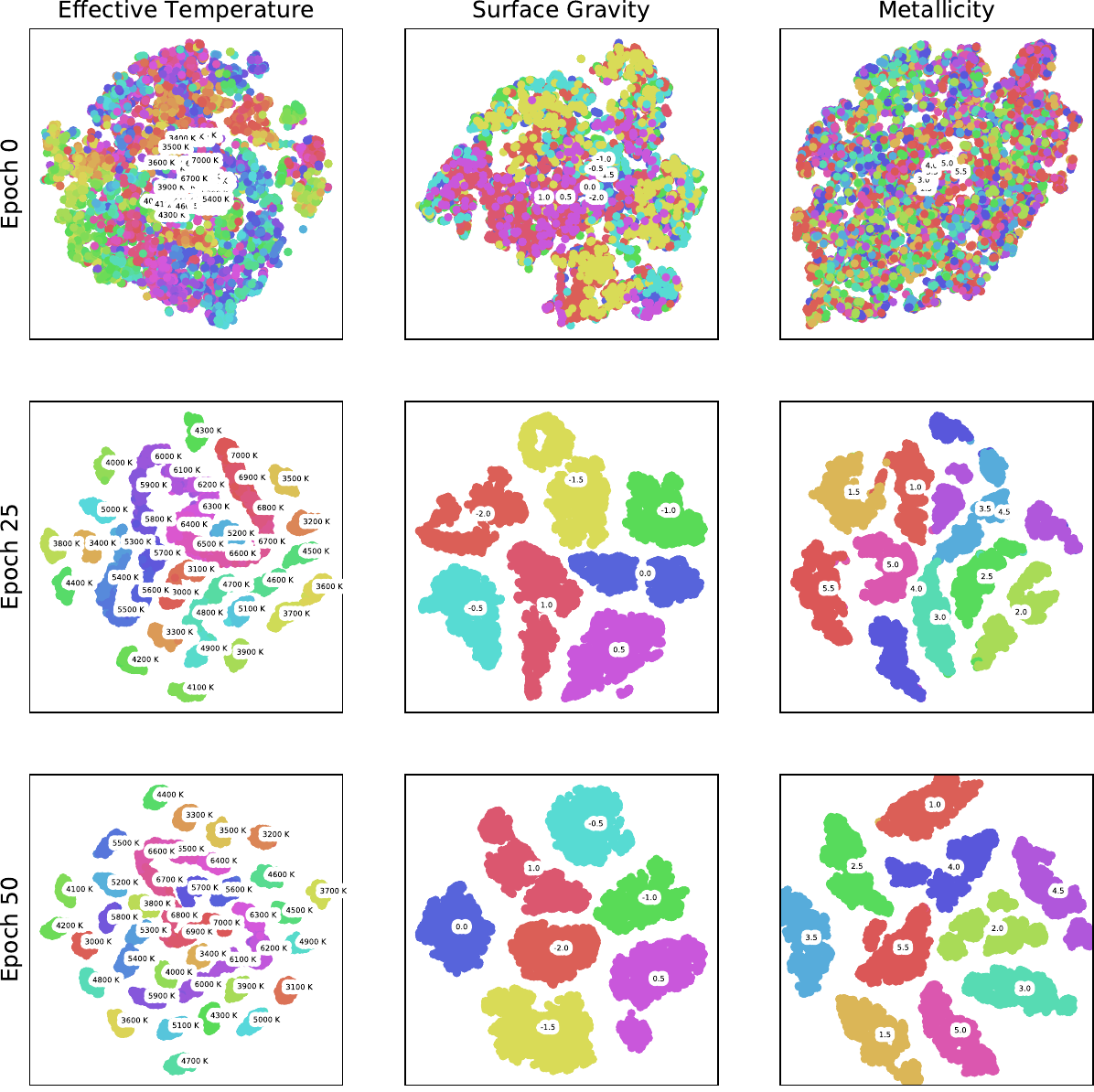}
    \caption{Visualization of the improved separability of stellar classes in the embedded feature space for the validation sample, facilitated by the triplet loss function.}
    \label{fig:triplet_val}
\end{figure*}

The efficacy of the triplet loss function in our $\rm{deep-REMAP}$ model is evident in Figure \ref{fig:triplet_val}, which showcases the improved separability of stellar classes in the learned feature space. This improvement is quantified by a marked increase in classification accuracy and a reduction in overlap between the parameter distributions of giants and dwarfs.

\section{Conclusions and Discussion}\label{sec:conclusions}

We have introduced $\rm{deep-REMAP}$, a pioneering neural network designed for the spectral analysis of 1D spectra, which we have utilized to parameterize MARVELS targets. Our model is the first of its kind to integrate an array of contemporary machine learning techniques including transfer learning, multi-task learning, temperature scaling, focal loss, triplet loss, stochastic weight averaging, cosine annealing-based learning rate, and probabilistic inference with an asymmetric cost function.

Our findings clearly demonstrate that, with carefully curated data augmentation and the application of transfer learning, our network can transcend the observational idiosyncrasies of specific spectroscopes. As a result, it is capable of estimating stellar parameters from real observed spectra with remarkable accuracy. Moreover, we have showcased how adopting a regression-as-classification approach allows us to capture the non-Gaussian distributions of the output parameters effectively. Furthermore, the incorporation of an embedding loss not only enhances the classification results but also significantly improves model interpretability.

For the first time, we present a methodology that predicts the parameters of stellar spectra with unprecedented accuracy, marking a paradigm shift in automated spectral analysis. This advancement paves the way for future surveys, providing a scalable and efficient tool for the rapid classification of an ever-growing number of stellar objects. The deep-REMAP network stands as a testament to the potential of machine learning in revolutionizing our understanding of the cosmos.





\appendix

\newpage
\section{Acklowledgements}
The author would like to thank Dr. Jian Ge and Mr. Kevin Willis, formerly at the University of Florida, for assistance with this project.

{\small
\bibliographystyle{IEEEtran}
\bibliography{neurips_2023_ref.bib} 

\begin{thebibliography}{10}
\providecommand{\url}[1]{#1}
\csname url@samestyle\endcsname
\providecommand{\newblock}{\relax}
\providecommand{\bibinfo}[2]{#2}
\providecommand{\BIBentrySTDinterwordspacing}{\spaceskip=0pt\relax}
\providecommand{\BIBentryALTinterwordstretchfactor}{4}
\providecommand{\BIBentryALTinterwordspacing}{\spaceskip=\fontdimen2\font plus
\BIBentryALTinterwordstretchfactor\fontdimen3\font minus \fontdimen4\font\relax}
\providecommand{\BIBforeignlanguage}[2]{{%
\expandafter\ifx\csname l@#1\endcsname\relax
\typeout{** WARNING: IEEEtran.bst: No hyphenation pattern has been}%
\typeout{** loaded for the language `#1'. Using the pattern for}%
\typeout{** the default language instead.}%
\else
\language=\csname l@#1\endcsname
\fi
#2}}
\providecommand{\BIBdecl}{\relax}
\BIBdecl

\bibitem{york2000sloan}
D.~G. York, J.~Adelman, J.~E. Anderson~Jr, S.~F. Anderson, J.~Annis, N.~A. Bahcall, J.~Bakken, R.~Barkhouser, S.~Bastian, E.~Berman \emph{et~al.}, ``The sloan digital sky survey: Technical summary,'' \emph{The Astronomical Journal}, vol. 120, no.~3, p. 1579, 2000.

\bibitem{eisenstein2011sdss}
D.~J. Eisenstein, D.~H. Weinberg, E.~Agol, H.~Aihara, C.~A. Prieto, S.~F. Anderson, J.~A. Arns, {\'E}.~Aubourg, S.~Bailey, E.~Balbinot \emph{et~al.}, ``Sdss-iii: Massive spectroscopic surveys of the distant universe, the milky way, and extra-solar planetary systems,'' \emph{The Astronomical Journal}, vol. 142, no.~3, p.~72, 2011.

\bibitem{segue}
B.~Yanny, C.~Rockosi, H.~J. Newberg, G.~R. Knapp, J.~K. Adelman-McCarthy, B.~Alcorn, S.~Allam, C.~A. Prieto, D.~An, K.~S. Anderson \emph{et~al.}, ``Segue: A spectroscopic survey of 240,000 stars with g= 14-20,'' \emph{The Astronomical Journal}, vol. 137, no.~5, p. 4377, 2009.

\bibitem{rave}
G.~Kordopatis, G.~Gilmore, M.~Steinmetz, C.~Boeche, G.~M. Seabroke, A.~Siebert, T.~Zwitter, J.~Binney, P.~De~Laverny, A.~Recio-Blanco \emph{et~al.}, ``The radial velocity experiment (rave): Fourth data release,'' \emph{The Astronomical Journal}, vol. 146, no.~5, p. 134, 2013.

\bibitem{boss}
K.~S. Dawson, D.~J. Schlegel, C.~P. Ahn, S.~F. Anderson, {\'E}.~Aubourg, S.~Bailey, R.~H. Barkhouser, J.~E. Bautista, A.~Beifiori, A.~A. Berlind \emph{et~al.}, ``The baryon oscillation spectroscopic survey of sdss-iii,'' \emph{The Astronomical Journal}, vol. 145, no.~1, p.~10, 2012.

\bibitem{lamost}
X.-Q. Cui, Y.-H. Zhao, Y.-Q. Chu, G.-P. Li, Q.~Li, L.-P. Zhang, H.-J. Su, Z.-Q. Yao, Y.-N. Wang, X.-Z. Xing \emph{et~al.}, ``The large sky area multi-object fiber spectroscopic telescope (lamost),'' \emph{Research in Astronomy and Astrophysics}, vol.~12, no.~9, p. 1197, 2012.

\bibitem{gaiaeso}
G.~Gilmore, S.~Randich, M.~Asplund, J.~Binney, P.~Bonifacio, J.~Drew, S.~Feltzing, A.~Ferguson, R.~Jeffries, G.~Micela \emph{et~al.}, ``The gaia-eso public spectroscopic survey,'' \emph{The Messenger}, vol. 147, pp. 25--31, 2012.

\bibitem{desi}
M.~Levi, C.~Bebek, T.~Beers, R.~Blum, R.~Cahn, D.~Eisenstein, B.~Flaugher, K.~Honscheid, R.~Kron, O.~Lahav \emph{et~al.}, ``The desi experiment, a whitepaper for snowmass 2013,'' \emph{arXiv preprint arXiv:1308.0847}, 2013.

\bibitem{lsst}
{\v{Z}}.~Ivezi{\'c}, S.~M. Kahn, J.~A. Tyson, B.~Abel, E.~Acosta, R.~Allsman, D.~Alonso, Y.~AlSayyad, S.~F. Anderson, J.~Andrew \emph{et~al.}, ``Lsst: from science drivers to reference design and anticipated data products,'' \emph{The Astrophysical Journal}, vol. 873, no.~2, p. 111, 2019.

\bibitem{mirkwood}
\BIBentryALTinterwordspacing
S.~Gilda, S.~Lower, and D.~Narayanan, ``mirkwood: Fast and accurate sed modeling using machine learning,'' \emph{The Astrophysical Journal}, vol. 916, no.~1, p.~43, jul 2021. [Online]. Available: \url{https://dx.doi.org/10.3847/1538-4357/ac0058}
\BIBentrySTDinterwordspacing

\bibitem{cfht}
\BIBentryALTinterwordspacing
S.~Gilda, S.~C. Draper, S.~Fabbro, W.~Mahoney, S.~Prunet, K.~Withington, M.~Wilson, Y.-S. Ting, and A.~Sheinis, ``{Uncertainty-aware learning for improvements in image quality of the Canada–France–Hawaii Telescope},'' \emph{Monthly Notices of the Royal Astronomical Society}, vol. 510, no.~1, pp. 870--902, 11 2021. [Online]. Available: \url{https://doi.org/10.1093/mnras/stab3243}
\BIBentrySTDinterwordspacing

\bibitem{neurips_cfht}
S.~{Gilda}, Y.-S. {Ting}, K.~{Withington}, M.~{Wilson}, S.~{Prunet}, W.~{Mahoney}, S.~{Fabbro}, S.~C. {Draper}, and A.~{Sheinis}, ``{Astronomical Image Quality Prediction based on Environmental and Telescope Operating Conditions},'' \emph{arXiv e-prints}, p. arXiv:2011.03132, Nov. 2020.

\bibitem{deepremap_abstract_aas}
S.~{Gilda}, J.~{Ge}, and {MARVELS}, ``{Parameterization of MARVELS Spectra Using Deep Learning},'' in \emph{American Astronomical Society Meeting Abstracts \#231}, ser. American Astronomical Society Meeting Abstracts, vol. 231, Jan. 2018, p. 349.02.

\bibitem{gamma_ray}
M.~{Dainotti}, V.~{Petrosian}, M.~{Bogdan}, B.~{Miasojedow}, S.~{Nagataki}, T.~{Hastie}, Z.~{Nuyngen}, S.~{Gilda}, X.~{Hernandez}, and D.~{Krol}, ``{Gamma-ray Bursts as distance indicators through a machine learning approach},'' \emph{arXiv e-prints}, p. arXiv:1907.05074, Jul. 2019.

\bibitem{valenti2005spectroscopic}
J.~A. Valenti and D.~A. Fischer, ``Spectroscopic properties of cool stars (spocs). i. 1040 f, g, and k dwarfs from keck, lick, and aat planet search programs,'' \emph{The Astrophysical Journal Supplement Series}, vol. 159, no.~1, p. 141, 2005.

\bibitem{equivalent_width}
\BIBentryALTinterwordspacing
S.~G. Sousa, \emph{ARES + MOOG: A Practical Overview of an Equivalent Width (EW) Method to Derive Stellar Parameters}.\hskip 1em plus 0.5em minus 0.4em\relax Cham: Springer International Publishing, 2014, pp. 297--310. [Online]. Available: \url{https://doi.org/10.1007/978-3-319-06956-2_26}
\BIBentrySTDinterwordspacing

\bibitem{ghezzi2014accurate}
L.~Ghezzi, L.~Dutra-Ferreira, D.~Lorenzo-Oliveira, G.~F.~P. De~Mello, B.~X. Santiago, N.~De~Lee, B.~L. Lee, L.~N. Da~Costa, M.~A. Maia, R.~L. Ogando \emph{et~al.}, ``Accurate atmospheric parameters at moderate resolution using spectral indices: Preliminary application to the marvels survey,'' \emph{The Astronomical Journal}, vol. 148, no.~6, p. 105, 2014.

\bibitem{mlp_1}
C.~A. Bailer-Jones, ``Stellar parameters from very low resolution spectra and medium band filters: teff, logg and [m/h] using neural networks,'' \emph{arXiv preprint astro-ph/0003071}, 2000.

\bibitem{mlp_2}
S.~Snider, C.~A. Prieto, T.~von Hippel, T.~C. Beers, C.~Sneden, Y.~Qu, and S.~Rossi, ``Three-dimensional spectral classification of low-metallicity stars using artificial neural networks,'' \emph{The Astrophysical Journal}, vol. 562, no.~1, p. 528, 2001.

\bibitem{mlp_3}
M.~Manteiga, D.~Ord{\'o}{\~n}ez, C.~Dafonte, and B.~Arcay, ``Anns and wavelets: A strategy for gaia rvs low s/n stellar spectra parameterization,'' \emph{Publications of the Astronomical Society of the Pacific}, vol. 122, no. 891, p. 608, 2010.

\bibitem{cannon}
M.~Ness, D.~W. Hogg, H.-W. Rix, A.~Y. Ho, and G.~Zasowski, ``The cannon: a data-driven approach to stellar label determination,'' \emph{The Astrophysical Journal}, vol. 808, no.~1, p.~16, 2015.

\bibitem{payne}
Y.-S. Ting, C.~Conroy, H.-W. Rix, and P.~Cargile, ``The payne: self-consistent ab initio fitting of stellar spectra,'' \emph{arXiv preprint arXiv:1804.01530}, 2018.

\bibitem{starnet}
S.~Bialek, S.~Fabbro, K.~A. Venn, N.~Kumar, T.~O'Briain, and K.~M. Yi, ``Deep learning analyses of synthetic spectral libraries with an application to the gaia-eso database,'' \emph{arXiv preprint arXiv:1911.02602}, 2019.

\bibitem{marvels_ge_intro}
J.~Ge, S.~Mahadevan, B.~Lee, X.~Wan, B.~Zhao, J.~van Eyken, S.~Kane, P.~Guo, E.~Ford, S.~Fleming \emph{et~al.}, ``The multi-object apo radial-velocity exoplanet large-area survey (marvels),'' in \emph{Extreme Solar Systems}, vol. 398, 2008, p. 449.

\bibitem{marvels_telescope}
J.~E. Gunn, W.~A. Siegmund, E.~J. Mannery, R.~E. Owen, C.~L. Hull, R.~F. Leger, L.~N. Carey, G.~R. Knapp, D.~G. York, W.~N. Boroski \emph{et~al.}, ``The 2.5 m telescope of the sloan digital sky survey,'' \emph{The Astronomical Journal}, vol. 131, no.~4, p. 2332, 2006.

\bibitem{marvels_ge_2}
\BIBentryALTinterwordspacing
J.~Ge, B.~Lee, N.~D. Lee, X.~Wan, J.~Groot, B.~Zhao, F.~Varosi, K.~Hanna, S.~Mahadevan, F.~Hearty, L.~Chang, J.~Liu, J.~van Eyken, J.~Wang, R.~Pais, Z.~Chen, A.~Shelden, and E.~Costello, ``{A new generation multi-object Doppler instrument for the SDSS-III Multi-object APO Radial Velocity Exoplanet Large-area Survey},'' in \emph{Techniques and Instrumentation for Detection of Exoplanets IV}, S.~B. Shaklan, Ed., vol. 7440, International Society for Optics and Photonics.\hskip 1em plus 0.5em minus 0.4em\relax SPIE, 2009, pp. 187 -- 196. [Online]. Available: \url{https://doi.org/10.1117/12.826651}
\BIBentrySTDinterwordspacing

\bibitem{grieves2017exploring}
N.~Grieves, J.~Ge, N.~Thomas, B.~Ma, S.~Sithajan, L.~Ghezzi, B.~Kimock, K.~Willis, N.~De~Lee, B.~Lee \emph{et~al.}, ``Exploring the brown dwarf desert: new substellar companions from the sdss-iii marvels survey,'' \emph{Monthly Notices of the Royal Astronomical Society}, vol. 467, no.~4, pp. 4264--4281, 2017.

\bibitem{Paegert_2015}
\BIBentryALTinterwordspacing
M.~Paegert, K.~G. Stassun, N.~D. Lee, J.~Pepper, S.~W. Fleming, T.~Sivarani, S.~Mahadevan, C.~E.~M. III, S.~Dhital, L.~Hebb, and J.~Ge, ``{TARGET} {SELECTION} {FOR} {THE} {SDSS}-{III} {MARVELS} {SURVEY},'' \emph{The Astronomical Journal}, vol. 149, no.~6, p. 186, may 2015. [Online]. Available: \url{https://doi.org/10.1088%2F0004-6256%2F149%2F6%2F186}
\BIBentrySTDinterwordspacing

\bibitem{grieves2018chemo}
N.~Grieves, J.~Ge, N.~Thomas, K.~Willis, B.~Ma, D.~Lorenzo-Oliveira, A.~Queiroz, L.~Ghezzi, C.~Chiappini, F.~Anders \emph{et~al.}, ``Chemo-kinematics of the milky way from the sdss-iii marvels survey,'' \emph{Monthly Notices of the Royal Astronomical Society}, vol. 481, no.~3, pp. 3244--3265, 2018.

\bibitem{cont_norm_wavelet}
R.~Zhao and A.~Luo, ``A novel method for continuum normalization of astronomical spectrum signals,'' \emph{Guang pu xue yu guang pu fen xi= Guang pu}, vol.~26, no.~3, pp. 587--590, 2006.

\bibitem{cont_norm_framelet}
H.~{Wang}, S.~{Li}, and A.~{Luo}, ``Astronomical spectral lines auto-extraction based on framelet transform,'' in \emph{2010 3rd International Congress on Image and Signal Processing}, vol.~3, Oct 2010, pp. 1045--1048.

\bibitem{cont_norm_wavelet_kalman}
\BIBentryALTinterwordspacing
S.~Gilda and Z.~Slepian, ``{Automatic Kalman-filter-based wavelet shrinkage denoising of 1D stellar spectra},'' \emph{Monthly Notices of the Royal Astronomical Society}, vol. 490, no.~4, pp. 5249--5269, 09 2019. [Online]. Available: \url{https://doi.org/10.1093/mnras/stz2577}
\BIBentrySTDinterwordspacing

\end{thebibliography}
}

\newpage
\begin{algorithm}
\caption{Continuum Normalization}
\label{algo:cont_norm}
\begin{algorithmic}[1]
\renewcommand{\algorithmicrequire}{\textbf{Input:}}
\renewcommand{\algorithmicensure}{\textbf{Output:}}
\REQUIRE a vector of spectrum flux measurements of length \emph{N}, $\textbf{Z} = \{z_n\}^{N}_{n=1}$.
\ENSURE the normalized spectrum flux vector of length \emph{N}, $\textbf{X} = \{x_n\}^{N}_{n=1}$.
\STATE Set $a$ as the maximum quantity of flux values in each bin that are allowed to be larger than the continuum.
\STATE Define the initial continuum estimate, $\textbf{C}$, as a vector of length $N$ with all elements set to 10 + max($\textbf{Z}$).
\FOR{bin count $v$ in range [2, 11]} 
\STATE Partition the index vector, $[1, N]$, into $v$ uniform bins. In each partition, determine the center element and place into vector $\textbf{H}$, $\textbf{H} = [N/v/2, \{[N/v/2, N]\}^{N}_{n=1+N/v}]$
\WHILE{the quantity of $\TRUE$ values in $\textbf{C} > \textbf{Z}$ is greater than $a$, }
\FOR{bin index $h$ in a random permuted vector of $\textbf{H}$} 
\STATE Create a subset of vector $\textbf{C}$ by duplicating indices $\textbf{H}$ of $\textbf{C}$ into vector $\textbf{E}$.
\IF{$\textbf{C}(h) > \textbf{Z}(h)$}
\STATE Lower the continuum estimate in bin $h$ by multiplying $\textbf{E}(h)$ by  0.999.
\IF{$v$ < 5}
\STATE Calculate the continuum estimate, $\textbf{C}$, by fitting a $v - 1$ degree polynomial to $(\textbf{H}, \textbf{E})$ and evaluate at range $[1, N]$.
\STATE To avoid undesired inflections at the continuum edges we perform linear interpolations to force flat edges in the endpoint bins. At the left edge we find the slope between the points $\textbf{C}(\textbf{H}(1))$ and $\textbf{C}(\textbf{H}(1) + 1)$, then extrapolate to replace the index range $[1, \textbf{H}(1) - 1]$. The right edge is altered in the same manner at the index range $[\textbf{H}(v) + 1, N]$.
\ELSE{}
\STATE Calculate the continuum estimate, $\textbf{C}$, by linear interpolation of $(\textbf{H}, \textbf{E})$ and evaluate at range $[1, N]$.
\ENDIF
\ENDIF
\ENDFOR
\ENDWHILE
\ENDFOR
\STATE Calculate the continuum estimate, $\textbf{C}$, by fitting a $6$ degree polynomial to $(\textbf{H}, \textbf{E})$ and evaluate at range $[1, N]$.
\STATE To avoid undesired inflections at the continuum edges we perform linear interpolations to force flat edges in the endpoint bins. At the left edge we find the slope between the points $\textbf{C}(\textbf{H}(1))$ and $\textbf{C}(\textbf{H}(1) + 1)$, then extrapolate to replace the index range $[1, \textbf{H}(1) - 1]$. The right edge is altered in the same manner at the index range $[\textbf{H}(v) + 1, N]$.
\STATE To smooth out the linear interpolations performed at the edges the continuum estimate is convolved with a gaussian kernel with sigma of 150.
\end{algorithmic} 
\end{algorithm}


\end{document}